\documentclass[aps,twocolumn,prl,showpacs,nofootinbib]{revtex4}
\usepackage{amsmath}
\usepackage{graphicx}
\usepackage{dcolumn}
\usepackage{bm}
\usepackage{amssymb}
\begin{document}
\title{Topology Changes and Quantum Phase Transition in
Spin-Chain System}
\author{Zhe Chang and Ping Wang}
\affiliation{Institute of High Energy Physics, Chinese Academy of
Sciences, P.O.Box 918(4), 100049 Beijing, China }
\email{changz@mail.ihep.ac.cn} \email{pwang@mail.ihep.ac.cn}

\begin{abstract}
The standard Landau-Ginzburg scenario of phase transition is broken
down for quantum phase transition. It is difficult to find an order
parameter to indicate different phases for quantum fluctuations.
Here, we suggest a topological description of the quantum phase
transition for the XY model. The ground states are identified as a
specialized $U(1)$ principal bundle on the base manifold $S^2$. And
then different first Chern numbers of $U(1)$ principal bundle on the
base manifold $S^2$ are associated to each phase of quantum
fluctuations. The particle-hole picture is used to parameterized the
ground states of the XY system. We show that a singularity of the
Chern number of the ground states occurs simultaneously with a
quantum phase transition.  The Chern number is a suitable
topological order of the quantum phase transition.

\end{abstract}
\pacs{75.10.Pq, 03.65.Vf, 03.65.Ud, 05.70.Jk}
 \maketitle
\section{introduction}
A classical phase transition brings about a sudden change of the
macroscopic properties of a many-body system while varying smoothly
temperature. The appearance of a singularity on canonical
thermodynamic function is a signature of a phase transition.
Furthermore, theorem of Lee and Yang relates the properties of the
zeros of the grandcanonical partition function in the complex
fugacity plane to singularity of the corresponding thermodynamic
function. In the Landau-Ginzburg scenario of phase transition, the
symmetry breaking described by order parameters indicates a phase
transition.  P. W. Anderson believed that the interaction of the
twin concepts of broken symmetry and of adiabatic continuity is the
logical core of many-body physics\cite{anderson}.

However, some systems found fails to fall in the  standard framework
of phase transition. Such as different fractional quantum Hall
states have the same symmetry but contains a completely new kind of
order (topological order). The classical statistical systems are
described by positive probability distribution functions of infinite
variables while fractional quantum Hall sates are described by their
ground state wave functions which are complex functions of infinite
variables. Quantum states contain a kind of order that is beyond
symmetry characterization\cite{niu-1,wen}.

For a many-body system at absolute zero temperature, all thermal
fluctuations are frozen out and quantum fluctuations retained only.
These microscopic quantum fluctuations can drive a macroscopic phase
transition. This phenomenon, known as quantum phase transition
(QPT), corresponds dramatic changes of ground states due to a small
variation of external parameters\cite{sachdev}. It is ascribed to
the interplay between different orderings associated to competing
terms in the Hamiltonian of the systems. QPT has attracted
considerable significance because of its association with condensed
matter physics and quantum information, though it is still difficult
to identify a proper parameter for indicating the symmetry breaking
in the QPT systems.

Recently, tools in theory of quantum information have been used to
characterize the critical points of QPTs.  The connections between
QPTs and quantum entanglement\cite{nielsen} was explored. Geometric
phase was used also as an indicator of QPTs\cite {carollo}. QPTs can
be studied systematically by means of differential geometry of
projective Hilbert space\cite{Zanardi1}. Geometric quantities of a
quantum many body system undergoing a QPT (in the thermodynamic
limit) display discontinuous features abruptly. In particular,
critical behavior corresponds properly to a singularity of the
metric equipped in parameter manifold of the ground states of the
quantum many-body systems. Moreover, the metric obeys scaling
behavior in the vicinity of a QPT. This differential geometry
approach stimulates  investigations of QPTs from a quite different
point of view.

In this Letter, within the framework of differential geometry
description of the QPTs, we develop a connection between a
singularity of the Chern number (a simple topological order)
associated with the ground states of the XY model and a transition
among different phases caused by quantum fluctuations. The ground
states of the XY system after rotating are parameterized by three
parameters, the anisotropy $\gamma$, rotating angle $\phi$ and the
magnetic field $\lambda$. For the case of $\gamma~ \in [0,\infty)$
(the other case of $\gamma~  \in (-\infty,0]$ can be discussed in
the same manner), the ground states can be identified as a specified
$U(1)$ principal bundle parameterized by $\lambda$ on the base
manifold $S^2_{\phi\gamma}$. We calculate the first Chern number (a
function of the magnetic field $\lambda$) for the ground states on
the base manifold $S^2_{\phi\gamma}$. It is found that different
Chern numbers of $U(1)$ principal bundle on the base manifold
$S^2_{\phi\gamma}$ are associated to each phase of quantum
fluctuations. The particle-hole picture of Lieb, Schultz and
Mattis\cite{lieb} is used to parameterized the ground states of the
XY system. We show that a singularity of the Chern numbers of the
ground states occur simultaneously with a quantum phase transition.
The Chern number is suggested as a suitable topological order of
QPTs.

\section{Differential geometric description}
QPTs behave distinctively from temperature-driven critical phenomena
as a consequence of competition between different parameters ,
$\Lambda\in \mathcal{M}$ (the parameter manifold), in the
Hamiltonian $H(\Lambda)$ of the system. For different Hamiltonians,
these parameters  describe different basic interactions,
respectively. QPTs take place for a parameter region where the
energy levels of the ground state and the excited state cross or
have an avoided crossing. The ground state $|\psi_0(\Lambda)\rangle$
gives a mapping of the parameter manifold $\mathcal{M}$ onto the
projective Hilbert space. In the projective Hilbert space, one can
define  naturally  a metric, the Fubini-Study metric\cite{nomizu}
\begin{equation}
ds^2=\langle d\psi_0|d\psi_0\rangle-\langle
d\psi_0|\psi_0\rangle\langle\psi_0|d\psi_0\rangle.
\end{equation}
In fact the  complex Hermitean tensor\cite{Zanardi1}
\begin{multline}
\mathcal{G_{\mu\nu}}=\langle\partial_\mu\psi_0|\partial_\nu\psi_0\rangle-
\langle\partial_\mu\psi_0|\psi_0\rangle\langle\psi_0|\partial_\nu\psi_0\rangle,\\\mu,\nu=1,...,
{\rm dim}\mathcal{M}~,
\end{multline}
can be introduced. The real component of $\mathcal{G}$ is just the
Fubini-Study metric expressed in the parameter space.

Generally, the definition of the Fubini-Study metric on the ground
states can be extended to arbitrary normalized eigenstates
$|\psi_n(\Lambda)\rangle$ of $H(\Lambda)$ with eigenvalue
$E_n(\Lambda)=\langle
\psi_n(\Lambda)|H(\Lambda)|\psi_n(\Lambda)\rangle$, and $ d^{2}
\left( \psi_n,\psi_n+d\psi_n \right)=\sum_{m\neq n}|\langle
\psi_m|d\psi_n\rangle |^{2}$. By using the differential of the
eigenstate equation $d([H-E_n]|\psi_n\rangle)=0$, we get
\begin{equation}
d^2(\psi_n,\psi_n+d\psi_n)=\sum_{m\neq
n}\frac{\langle\psi_m|dH|\psi_n\rangle\langle
\psi_n|dH|\psi_m\rangle}{(E_m-E_n)^2}.
\end{equation}
For the ground state, we can rewrite the complex Hermitean tensor
$\mathcal{G}$ as
\begin{equation}
\mathcal{G}=\sum_{m\neq 0}\frac{\langle
\psi_m|dH|\psi_0\rangle\langle\psi_0|dH|\psi\rangle}{(E_m-E_0)^2},\label{g}
\end{equation}
where $E_0$ and $E_m$ are the energy of the ground state and excited
state respectively, and the summation runs over all excited states
$|\psi_m\rangle ~(m=1,2...)$ of the system.

In the thermodynamic limit, QPTs take place where the energy gap
between the ground state and the first excited state is vanishing
for some specific parameters region. This is reflected in (\ref{g})
where the denominator is vanishing and $\mathcal{G}$ is divergent,
resultantly.

The real component of $\mathcal{G}$ induces a Riemannian metric in
parameter manifold which behaves singularly while parameters
approaching critical points. This property of induced metric in the
parameter manifold has been investigated in detail for various
models\cite{Zanardi1}. Here we focus on imaginary component of the
complex Hermitean tensor $\mathcal{G_{\mu\nu}}$,
\begin{equation}
F_{\mu\nu}=\langle\partial_\mu\psi_0|\partial_\nu\psi_0\rangle\\
-\langle\partial_\nu\psi_0|\partial_\mu\psi_0\rangle~.\label{Gmunv}
\end{equation}
$F_{\mu\nu}$ is the Berry curvature two-form\cite{berry}. The ground
state is a $U(1)$ principal bundle on the parameter manifold
${\mathcal M}$. The principal bundle can be classified by
topological parameters. This, in fact, gives an explicit
clarification of the figurations of ground states for a kind of
general systems. Crossing of phase boundary on the parameter space
corresponds to a singularity of the curvature two form $F_{\mu\nu}$.
It is natural to expect that different winding numbers of the $U(1)$
principal bundle can be associated to different quantum phases.

\section{the ground state of the XY model}
In what follows, we present our arguments through the familiar XY
model.  The quantum XY model is a one dimensional spin-1/2 chain
with nearest-neighbor interaction. The Hamiltonian of the model is
described by

\begin{equation}
H(\gamma,\lambda)=-\frac{1}{2}\sum^N_{j=1}\left(\frac{1+\gamma}{2}\sigma^x_j\sigma^x_{j+1}\\
+\frac{1-\gamma}{2}\sigma^y_j\sigma^y_{j+1}+\lambda\sigma^z_j\right)~,\label{H}
\end{equation}
where $\sigma^x_j$, $\sigma^y_j$, $\sigma^z_j$ represent Pauli
matrices at $j$-th lattice site. The parameter $\gamma$ denotes the
anisotropy in the in-plane nearest-neighbor interaction, and
$\lambda$ is the transverse field applied in $z$ direction.

Critical behavior occurs at magnetic field $\lambda=\pm1$ for any
value of $\gamma$ and $\gamma=0$ for the value of $\lambda\neq\pm1$.
The eigenstate and eigenenergy can be exactly calculated by means of
diagonalizing the Hamiltonian\cite{lieb}. The most important step is
the Jordan-Wigner transformation which maps spin-1/2 degrees of
freedom to spinless fermions.

For the purpose of investigating topology of the system, we
introduce another Hamiltonian\cite{carollo} $H
(\phi,\gamma,\lambda)\equiv R(\phi) H(\gamma,\lambda)
R^\dagger(\phi)$ of the model by rotating
$R(\phi)=\prod^N_{j=1}\exp(i\phi\sigma^z_j/2)$ every spin in system
around the $z$ direction with an angle $\phi\in [0,\pi)$. The
Hamiltonian can be diagonalized by a standard procedure. By making
use of the Jordan-Wigner transformation
\begin{equation}
a_l=\left(\prod_{m<l}\sigma^z_m\right)\frac{\sigma^x_l+i\sigma^y_l}{2}~,
\end{equation}
one converts the spin operators into fermionic operators
\begin{multline}
H(\phi,\gamma,\lambda)=\\-\frac{1}{2}\sum^{N-1}_{j=1}(a^\dagger_{j+1}a_j+a^\dagger_j
a_{j+1}-\gamma e^{2i\phi}a_j a_{j+1}+\gamma e^{-2i\phi}a^\dagger_j
a^\dagger_{j+1})\\+\lambda\sum^N_{j=1}(a^\dagger_j
a_j)+\frac{1}{2}\alpha(a^\dagger_1 a_N-a^\dagger_N
a_1)+\frac{1}{2}\alpha\gamma(e^{-2i\phi}a^\dagger_1
a^\dagger_N-e^{2i\phi}a_N a_1)\\-\frac{N}{2}\lambda,
\end{multline}
where $\alpha=\prod^{N-1}_{j=1}(1-2a^\dagger_j a_j)$ and satisfies
$\alpha^2=1$, as well as  $[H,\alpha]=0$. $H$ and $\alpha$ are
simultaneously diagonalizable with eigenvalue of $\alpha=\pm 1$. For
large system, we may neglect boundary terms. The Fourier
transformation gives

\begin{equation}
d_k=\frac{1}{\sqrt{N}}\sum^N_{l=1}a_le^{-i2\pi lk/N}~.
\end{equation}
Making use of the Bogoliubov transformation

\begin{equation}\begin{array}{rcl}
c_k&=&d_k\cos\frac{\theta_k}{2}-id^\dagger_{-k}e^{-2i\phi}\sin\frac{\theta_k}{2}~,\\[0.2cm]
c_k^\dagger&=&d_k^\dagger\cos\frac{\theta_k}{2}+id_{-k}e^{2i\phi}\sin\frac{\theta_k}{2}~,\\[0.2cm]
c_{-k}&=&d_{-k}\cos\frac{\theta_k}{2}+id^\dagger_k
e^{-2i\phi}\sin\frac{\theta_k}{2}~,\\[0.2cm]
c^\dagger_{-k}&=&d^\dagger_{-k}\cos\frac{\theta_k}{2}-id_k
e^{2i\phi}\sin\frac{\theta_k}{2}~,
\end{array}
\end{equation}
we obtain the  Hamiltonian  in a diagonal form

\begin{equation}\label{hamiltonian}
H(\phi,\gamma,\lambda)=\sum_{k=-[N/2]}^{[N/2]}\Lambda_k\left(c^\dagger_kc_k-\frac{1}{2}\right)~,
\end{equation}
where $\Lambda_k$ is the dispersion relation of the collective
excitation mode $c^\dagger_k$,

\begin{equation}\label{dispersion}
\Lambda_k=\pm\sqrt{\left(\cos\frac{2\pi
k}{N}-\lambda\right)^2+\gamma^2\sin^2\frac{2\pi k}{N}}~.
\end{equation}
The angle $\theta_k$ is determined by
$\cos\theta_k=(\lambda-\cos\frac{2\pi k}{N})/|\Lambda_k|$ and the
sign of $\Lambda_k$ is arbitrary. It should be noticed that the
adoption of the sign of the dispersion relation does not affect the
diagonalization of the Hamiltonian, but makes ambiguity in defining
the ground state of the system.

To solve the ambiguity, we first focus on the case of $\gamma=0$.
Note that the model is isotropic and diagonalized without performing
a Bogoliubov transformation. The Hamiltonian is
$H(\phi,0,\lambda)=\displaystyle\sum_{k=-[N/2]}^{[N/2]}\left(\lambda-\cos\frac{2\pi
k}{N}\right)d^\dagger_k d_k$. The ground state is of the form,
\begin{equation}
\begin{array}{rcl}
|\psi_0\rangle&=&
\displaystyle\prod^{-(k_T+1)}_{k=-[N/2]}|0\rangle\otimes\prod_{k=-k_T}^{k_T}|1\rangle_k
\otimes \prod_{k=k_T+1}^{[N/2]}|0\rangle~,\\[0.8cm]
k_T&=&\left\{\begin{array}{l}\left[\frac{N}{2\pi}\arccos\lambda\right]~,~~~~|\lambda|\leq
1~,\\[0.2cm]0~,~~~~~~~~~~~~~~~~~~~~|\lambda|>1~. \end{array}\right.
\end{array}\end{equation}
In the particle-hole picture of Lieb, Schultz and Mattis\cite{lieb},
the Fermi energy of ground states for the isotropic XY model is
given by $\frac{2\pi k_T}{N}=\arccos\lambda$.

Another important case, where the sign of the dispersion relation is
determined explicitly, is of $\gamma=\sqrt{1-\lambda^2}$. The
diagonalized Hamiltonian is of the form
\begin{multline}
H(\phi,\sqrt{1-\lambda^2},\lambda)=\displaystyle\sum_{k=-[N/2]}^{[N/2]}\sqrt{1-\gamma^2}\\\times\left(\frac{\lambda}
{1-\gamma^2}-\cos\frac{2\pi k}{N}\right)\left(c^\dagger_k c_k
-\frac{1}{2}\right)~.
\end{multline}
The ground state $|\psi_0\rangle$ can be constructed as

\begin{equation} \label{ground state1}
|\psi_0\rangle=\displaystyle\prod_{k=-[N/2]}^{[N/2]}
\left(\cos\frac{\theta_k}{2}|0\rangle_k|0\rangle_{-k}
+ie^{-2i\phi}\sin\frac{\theta_k}{2}|1\rangle_k|1\rangle_{-k}\right)~,
\end{equation}
where $|0\rangle_k$ denotes vacuum of $d_k$, and
$|1\rangle_k=d_k^\dagger|0\rangle_k$. \\
Generally, we can rewrite the Hamiltonian as
\begin{multline}
H(\phi,\gamma,\lambda)=\sum^{-(k_T+1)}_{k=-
[N/2]}\vert\Lambda_k\vert\left(c^\dagger_k
c_k-\frac{1}{2}\right)\\-\sum_{k=-k_T}^{k_T}\vert\Lambda_k\vert\left(c^\dagger_k
c_k-\frac{1}{2}\right)
+\sum^{[N/2]}_{k=k_T+1}\vert\Lambda_k\vert\left(c^\dagger_k
c_k-\frac{1}{2}\right)~,
\end{multline}
where we have used the notation
$$k_T\equiv\left\{\begin{array}{l}
\left[\frac{N}{2\pi}\arccos\frac{\lambda}{1-\gamma^2}\right]~,~~~~~{\rm
for~} \vert\frac{\lambda}{1-\gamma^2}
\vert\leq 1\\[0.2cm]
0~,~~~~~~~~~~~~~~~~~~~~~~~~~{\rm otherwise}~.
\end{array}\right.$$
$c_k^\dagger$ creates a hole if $0\leq\frac{2\pi
|k|}{N}\leq\arccos\frac{\lambda}{1-\gamma^2}$ for $\vert
\frac{\lambda}{1-\gamma^2 }\vert\leq 1$. But $c_k^\dagger$ creates a
particle if $\pi\geq\frac{2\pi
|k|}{N}>\arccos\frac{\lambda}{1-\gamma^2}$ for $\vert
\frac{\lambda}{1-\gamma^2 }\vert\leq 1$. In the case of $\vert
\frac{\lambda}{1-\gamma^2 }\vert>1$, no hole excitation mode exists.
The corresponding ground state of the Hamiltonian
$H(\phi,\gamma,\phi)$ is of the form

\begin{equation}\begin{array}{rcl}\label{ground state1}
|\psi_0\rangle&=&\displaystyle\prod_{k=-[N/2]}^{-(k_T+1)}
\left(\cos\frac{\theta_k}{2}|0\rangle_k|0\rangle_{-k}
+ie^{-2i\phi}\sin\frac{\theta_k}{2}|1\rangle_k|1\rangle_{-k}\right)\\[0.25cm]
& &\otimes\displaystyle\prod_{k=-k_T}^{k_T}
\left(\cos\frac{\theta_k}{2}|1\rangle_k|1\rangle_{-k}
-ie^{2i\phi}\sin\frac{\theta_k}{2}|0\rangle_k|0\rangle_{-k}\right)\\[0.25cm]
& &\otimes\displaystyle\prod^{[N/2]}_{k=k_T+1}
\left(\cos\frac{\theta_k}{2}|0\rangle_k|0\rangle_{-k}
+ie^{-2i\phi}\sin\frac{\theta_k}{2}|1\rangle_k|1\rangle_{-k}\right)~.
\end{array}
\end{equation}
In the case of $\lambda=0$, the Fermi energy reduces to $\pi/2$.
This is just the scenario described by Lieb, Schultz and
Mattis\cite{lieb}.

\section{topological order and quantum phase transition}
There are three parameters $\lambda$, $\gamma$ and $\phi$ in the
Projective Hilbert space $\mathcal {PH}$ of the ground states. We
note the fact that the Hamiltonian $H(\phi,\gamma,\lambda)$ is $\pi$
periodic in $\phi$. The projective Hilbert space $\mathcal {PH}$ can
be viewed as a $U(1)$ principal bundle parameterized  by  $\lambda$
on the base manifold $S^2_{\phi\gamma}$ ($\gamma\in
[0,\infty),~\phi\in [0,\pi)$).

The curvature tensor of the principal bundle on the base manifold
$S^2_{\phi\gamma}$ is of the form

\begin{multline}
F_{\phi\gamma }=\langle\frac{\partial
\psi_0}{\partial\phi}\vert\frac{\partial
\psi_0}{\partial\gamma}\rangle-\langle\frac{\partial
\psi_0}{\partial\gamma}\vert\frac{\partial
\psi_0}{\partial\phi}\rangle\\
=\frac{2i}{\pi}(\int_{2\pi k_T/N}^\pi
d\alpha\frac{\gamma\sin^2\alpha(\lambda-cos\alpha)}{[(\cos\alpha-\lambda)^2+\gamma^2\sin^2\alpha]^{3/2}}\\
-\displaystyle\int^{2\pi k_T/N}_0
d\alpha\frac{\gamma\sin^2\alpha(\lambda-cos\alpha)}{[(\cos\alpha-\lambda)^2+\gamma^2\sin^2\alpha]^{3/2}})~.
\end{multline}
Here the sum in the above equation has been
replaced by integration, as in the following, we will discuss the
topological changes caused by QPT's in the thermodynamic limit.

The first Chern number\cite{choquet} of the ground states on the
base manifold $S^2_{\phi\gamma}$ is\cite{pwang}

\begin{equation}\label{chern2}
C_1(\lambda)=\displaystyle \frac{i}{2\pi}\int^\pi_0
d\phi\int^{\infty}_{0} d\gamma F_{\phi\gamma}
=\left\{\begin{array}{l} -1,~~~~~~~{\rm for}~0\leq\lambda<1~,\\
-2,~~~~~~~{\rm for}~\lambda=1.\\
 ~0~,~~~~~~~{\rm
for}~\lambda>1.\end{array} \right.
\end{equation}
According to the result (\ref{chern2}) on the first Chern numbers of
the ground states, we can draw a phase diagram for the XY model. It
is clear to see that it is the same with the QPT diagram. Thus, the
Chern number can be really employed as a tool in describing the
QPTs.

\section{conclusions and perspectives}
The result obtained in the thermodynamic limit shows  clearly  that
the QPTs can be associated with a change in the topology of the
ground states. The topological description of the QPT's presented
here shares remarkable similarity to the topological order picture
of the integer quantum Hall effect\cite{niu-1,niu}, where Hall
conductance is explained as a topological invariant. Further
investigations of the similarities between QPT's in the spin chain
and quantum Hall effect may be interesting and we hope they will
benefit from each other.

Recently, quantum entanglement is intensively studied in quantum
many-body systems\cite{nielsen}. The entanglement changes
drastically while a parameter varying smoothly and the system
crossing through critical points. This phenomena  promotes the
understanding of the relation between the entanglement and QPTs.  It
also has been suggested that the classical phase transition might
has its deep origin in topological change of configuration
space\cite{casetti}. It is nature to wish a same topological
perspective on the phase transition of both classical and quantum
many body systems. Further investigations on the relation between
them is worthwhile.

\section{acknowledgements}
We would like to thank T. Chen and Y. Yu for cooperation at the
early stage of the paper. It is indebted to X. Li and S. J. Qin for
useful discussion on topology of principal bundle and Fermion
picture of lattice models. The work was supported by the NSF of
China under Grant No. 10575106.


\begin{thebibliography}{99}
\bibitem{anderson} P. W. Anderson, {\em Basic Notions of Condensed Matter
Physics}, The Benjamin-Cummings Publishing Company, Inc. London,
1984.
\bibitem{niu-1} Q. Niu, D. J. Thouless and Y. S. Wu, Phys. Rev. B{\bf 31},
3372 (1985).
\bibitem{wen} X. G. Wen and Q. Niu, Phys. Rev. B{\bf 41}, 9377
(1990).
\bibitem{sachdev} S. Sachdev, {\em Quantum Phase Transitions}, Cambridge
University Press, Cambridge, England, 1999.
\bibitem{nielsen} A. Osterloh, L. Amico, G. Falci, and R. Fazio, Nature
{\bf 416}, 608 (2002); T. J. Osborne and M. A. Nielsen, Phys. Rev.
A{\bf 66}, 032110 (2002); G. Vidal, J. I. Latorre, E. Rico, and A.
Kitaev, Phys. Rev. Lett. {\bf 90}, 227902 (2003); L. A. Wu, M. S.
Sarandy, and D. A. Lidar, Phys. Rev. Lett. {\bf 93}, 250404 (2004);
T. R. de Oliveira, G. Rigolin, M. C. de Oliveira, and E. Miranda,
Phys. Rev. Lett. {\bf 97}, 170401 (2006).
\bibitem{carollo} A. C. M. Carollo
and J. K. Pachos, Phys. Rev. Lett. {\bf 95}, 157203 (2005); S. L.
Zhu, Phys. Rev. Lett. {\bf 96}, 077206 (2006); A. Hamma,
quant-ph/0602091.
\bibitem{Zanardi1} P. Zanardi, P. Giorda, and M. Cozzini,
Phys. Rev. Lett. {\bf 99}, 100603 (2007) ; L. CamposVenuti and P.
Zanardi,
 Phys. Rev. Lett. {\bf 99}, 095701 (2007).
\bibitem{lieb} E. Lieb, T. Schultz, and D. Mattis, Ann. Phys. (N.Y.) {\bf 16},
407 (1961).
\bibitem{nomizu} S. Kobayashi and K. Nomizu, {\em Foundations of Differential
Geometry}, Interscience, New York, 1969; J. Anandan and Y. Aharonov,
Phys. Rev. Lett. {\bf 65}, 1697 (1990).
\bibitem{berry} M.V. Berry, Proc. R. Soc. London A{\bf 392}, 45
(1984); Y. Aharonov and J. Anandan, Phys. Rev. Lett. {\bf 58}, 1593
(1987); {\em Geometric Phases in Physics}, edited by A. Shapere and
F. Wilczek, World Scientific, Singapore, 1989.
\bibitem{choquet} Y. Choquet-Bruhat, C. DeWitt-Morette, and M. Dillard-Bleick,
{\em Analysis, Manifolds and Physics}, North-Holland, Amsterdam,
1982.
\bibitem{pwang} The numerical computation of the Chern numbers shows
amount of deviation from the data near critical points because of
singularity behavior of the curvature.
\bibitem{niu} D. J. Thouless, M. Kohmoto, M. P. Nightingale and M. den Nijs, Phys. Rev.
Lett. {\bf 49}, 405 (1982); J. E. Avron and R. Seiler, Phys. Rev.
Lett. {\bf 54}, 259 (1985).
\bibitem{casetti} L. Casetti, M. Pettini, and E. G. D. Cohen, Phys. Rep. {\bf 337}, 237 (2000);
 L. Casetti, M. Pettini, and E. G. D. Cohen, J. Stat.
Phys. {\bf 111}, 1091 (2003); L. Caiani, L. Casetti, C. Clementi and
M. Pettini, Phys. Rev. Lett. {\bf 79}, 4361 (1997); R. Franzosi, M.
Pettini, and L. Spinelli, Phys. Rev. Lett. {\bf 84}, 2774 (2000); M.
Kastner, Rev. Mod. Phys. {\bf 80}, 167 (2008).
\end{thebibliography}
\end{document}